\documentclass[useAMS,usenatbib]{mn2e}
\usepackage{graphicx}

\title[1001 Nova Outbursts]{A Thousand and One Nova Outbursts}

\author[N. Epelstain O. Yaron A. Kovetz and D. Prialnik]{Noya Epelstain\thanks{E-mail: noyaepel@post.tau.ac.il}, 
Ofer Yaron\thanks{E-mail: oferya@post.tau.ac.il}, 
Attay Kovetz\thanks{E-mail: attay@etoile.tau.ac.il} 
and 
Dina Prialnik\thanks{E-mail: dina@planet.tau.ac.il}
\\ 
Department of Geophysics and Planetary Sciences, Sackler Faculty of Exact Sciences \\
Tel Aviv University, Tel Aviv 69978, Israel
}
\begin{document}

\date{Accepted . Received ; in original form .}

\pagerange{\pageref{firstpage}--\pageref{lastpage}} \pubyear{2006}

\maketitle

\label{firstpage}

\begin{abstract}

A full nova cycle includes 
mass accretion, thermonuclear runaway resulting in outburst and mass loss, and finally, decline.  Resumed accretion 
starts a new cycle, leading to another outburst. 
Multicycle nova evolution models have been calculated over the past twenty years, the number being 
limited by numerical constraints. Here we present a long-term evolution code 
that enables a continuous calculation through an unlimited number of nova cycles for an unlimited evolution time, even up
to $ 1.5 \times 10 ^ {10} $~yr. 
Starting with two sets of the three independent nova parameters --- the white dwarf mass, the temperature of its isothermal core,
and the rate of mass transfer on to it --- we have followed the evolution of two models, 
with initial masses of 1~$M_\odot $ and 0.65~$M_\odot $ through over 1000 and over 3000 cycles, respectively. 
The accretion rate was assumed constant throughout each calculation: $10^{-11}~M_\odot$~yr$^{-1}$ for the 1~$M_\odot$ white dwarf, and
$10^{-9}~M_\odot$~yr$^{-1}$ for the 0.65~$M_\odot$ one. The initial temperatures were taken to be relatively high: 
$30\times10^6$~K and $50\times10^6$~K, respectively, as they are likely to be at the onset of the outburst phase.
The results show that although on the short-term consecutive outbursts are almost identical, on the long-term scale the characteristics
change. This is mainly due to the changing core temperature, which decreases very similarly to that of a cooling white dwarf 
for a time, but at a slower rate thereafter. As the white dwarf's mass continually decreases, since both models lose more mass 
than they accrete, the central pressure decreases accordingly. The outbursts on the massive white dwarf change gradually 
from fast to moderately fast, and the other characteristics (velocity, abundance ratios, isotopic ratios) change, too. 
Very slowly, a steady state is reached, where all characteristics, both in quiescence and in outburst, remain almost 
constant. For the less massive white dwarf accreting at a high rate, outbursts are similar throughout the evolution.
 
\end{abstract}


\section{Introduction}
\label{intro}

It is now twenty years since the publication of the first numerical simulation of a full nova cycle (Prialnik, 1986). During this time 
modelling of nova outbursts has advanced significantly and calculations have extended to continuous evolution  thorough a series of
several cycles (e.g., Shara et al., 1993). Extensive parameter studies of the classical nova phenomenon have been 
carried out by Prialnik and Kovetz (1995) and Kovetz and Prialnik (1997), and more recently by Yaron et al. (2005). 
In parallel, 
computation capability has increased tremendously and thus the computing time required to simulate a full nova cycle 
has dropped from months to minutes.

Models have shown --- and observations have confirmed --- that the characteristics of a nova outburst are determined 
by three independent parameters (Schwartzman et al., 1994; Prialnik and Kovetz, 1994; Prialnik, 1995). 
On the theoretical side, these are the mass and core temperature
(or luminosity) of the white dwarf (WD), which accretes mass from a binary companion and eventually ignites it explosively and
ejects it back into space, and the rate of accretion on to the WD. Changing these parameters independently over
the entire ranges allowed by stellar structure theory reproduces the wide range of observed characteristics of novae (Yaron et al.,
loc. cit.). So far, therefore, the theory of nova outbursts has proved extremely successful. However, the three free parameters
are not truly independent. They are linked by the long-term evolution of the binary system, which changes all of them simultaneously.
Thus the question remains whether the entire 3-dimensional parameter space is accessible to nova binary systems. 
Is it possible for the WD to cool to low temperatures despite repeated outbursts and loss or gain of mass? Does the WD mass remain
the exclusive structural parameter, or is the structure affected on the long-term scale by the events occurring at the surface? 
Such questions can only be answered by following the evolution of an accreting WD continuously, not through one or a few nova
cycles, but through a thousand such cycles and more. This is the purpose of the present paper. 
Long-term unabridged evolutionary calculations through repeated nova outburst cycles require some changes in the evolution 
code used in previous calculations, which will be described in Section~\ref{method}. 
The results of two computation runs are described and analysed in Section~\ref{ncyc}, both from the point of view of the 
outbursts and of the WD on top of which they occur. A brief discussion, summary and conclusions are given in Section~\ref{conclusions}.

\section{Long-term evolutionary calculations}
\label{method}

The hydrodynamic Lagrangian stellar evolution code used for the calculations presented here was described in some detail by 
Prialnik and Kovetz (1995). It includes OPAL opacities, a nuclear reactions network among 40 heavy element isotopes up to $^{31}$P,
convection according to the mixing-length theory, diffusion for all elements, accretional heating, and a mass-loss algorithm
that applies a steady, optically thick supersonic wind solution. 
We note, in particular, that the dynamical phases are calculated by solving the equation of
motion along with the energy balance equation (rather than imposing hydrostatic equilibrium). Mass 
loss is calculated continuously, according to the mass loss rate $\dot M_{m-l}$ derived from the optically thick wind solution. 
At each time step $\delta t$, an amount $\dot M_{m-l}\delta t$ is subtracted from the outermost mass shell $\Delta m_S$. 
Time steps are constrained during this phase by the requirement $\dot M_{m-l}\delta t < \Delta m_S$. Whenever $\Delta m_S$ becomes
very small, it is merged with the underlying mass shell. In a similar manner, during the accretion phase, an amount 
$\dot M_{acc}\delta t$ is added to the outermost mass shell, and whenever $\Delta m_S$ exceeds a prescribed value, the shell is
divided into two mass shells.

Although thousands of cycles are computed in the present study, each nova cycle, including all evolutionary phases and involving
a large network of nuclear reactions, is calculated accurately, with no short-cuts or simplifications.
A detailed description of the changing physical parameters during the evolution of a full nova
outburst was given by Prialnik (1986). Here we refrain from dealing with any single outburst cycle in detail, but focus on the
trends of change of typical characteristics and on long-term evolutionary effects. 

The spherical grid adopted for the numerical solution, which
includes the entire WD, is fixed, except for the outermost grid shells, which may grow or shrink, and be divided or merged,
during the accretion and mass-loss episodes of the nova cycle.
The fixed grid is, however, by no means an equally-spaced one. The
masses of shells vary by many orders of magnitude, according to the gradients of various physical properties (such as temperature,
density, element abundances) that are expected to develop during evolution. Thus, the spatial grid must be prepared 
in advance in anticipation of the evolutionary course of the stellar model. In particular, the outer layers of the WD that are expected
to take part in the various outburst processes (diffusion, convection, nuclear burning, acceleration, expansion, ejection, 
contraction, and so forth) must be finely zoned. Since, as a rule, each outburst erodes the WD of some mass 
(in addition to the accreted mass), the grid is set
in advance for a prescribed number of nova cycles. 
The larger the number of cycles, the larger the grid, 
and consequently, the larger the computation time.
This was the procedure adopted in former multicycle calculations.
Thus the number of cycles was restricted in the original form of the code by the fixed spacial grid of the numerical scheme.

The long-term evolution code should allow an unlimited number of cycles. 
A new algorithm is therefore devised whose purpose is to rezone the grid at the beginning of each nova cycle, thereby preparing it
for the upcoming evolutionary stages of that cycle. This numerical process, which involves interpolation on a previous grid, 
is designed in such 
a way as to conserve energy and mass of each nuclear species. Hydrostatic equilibrium (which prevails between the dynamical 
phases of the outburst itself) is preserved as well.

One of the problems that have to be overcome is the determination of the optimal mass shell size for the outer part of the grid. 
Since consecutive nova
outbursts are very similar (almost identical), this is done at each cycle based on the characteristics of
the previous one, namely, the amount of accreted mass and ejected mass, and the depth of the burning shell and convective zone.
Another problem is to determine the depth of the finely zoned region and the rate at which to gradually increase
shell masses until they eventually match the coarsely zoned bulk of the star. This is achieved by a geometrically increasing series,
with a typical factor of 1.25.
Since matrix inversion is required at each iteration of every time-step of the numerical calculation, a compromise must
be reached between computing time and resolution (or accuracy). Typically, the grid included between 200 and 400 mass 
shells for the entire configuration and each continuous --- hands-off --- run took several weeks of CPU time on a fast PC.

Finally, the code is capable of correcting itself in case of failure, by going back to an earlier stage and continuing from that
stage with adjusted numerical parameters, such as length of time-step, maximal number of iterations allowed for
achieving convergence, and so forth. Occasionally, a couple of times per thousand cycles, it does happen that the code goes astray,
but it recovers without intervention. Given the complexity of the input physics, in particular the need to interpolate
among a large number of opacity tables, this should not be surprising. Such stray points have been removed from the results.

A run spanning thousands of nova outburst cycles produces a prohibitive amount of data, 
which must be carefully selected, divided among a large number
of files, stored, and analysed, mainly by graphical means.
A great deal of information is necessarily lost in the process and can only be retrieved by repeating the calculation.

\section[]{Evolution through repeated nova cycles}
\label{ncyc}

Two evolution runs were carried out, starting from two different initial parameter combinations taken from the
extended grid of nova models studied by Yaron et al. (2005). We chose a relatively massive WD, of relatively high temperature, 
accreting at a low rate (hereafter, Model A), and a hot, low-mass WD, accreting at a relatively high rate (hereafter, Model B). 
According to Prialnik and Kovetz (1995) and
Yaron et al. (2005), the outburst characteristics of these models differ considerably, although in both cases they were found
to be well within the observed range of nova properties.

\begin{figure}
\centering
\includegraphics[width=90mm]{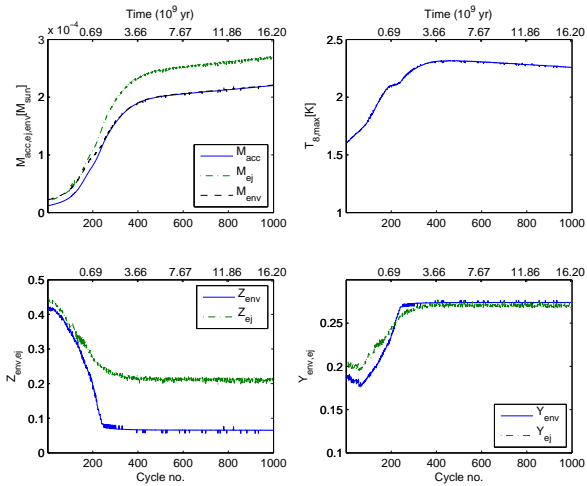}
\caption{Evolution of Model A ($ M_{WD} = 1 M_\odot $, $\dot{M}=10^{-11} M_\odot$~yr$^{-1}$ and  $T_{WD}=30\times10^6$~K): 
{\it top left:} accreted/ejected/envelope mass; {\it top right:} maximal temperature attained at outburst; 
{\it bottom left:} ejected/envelope metallicity; {\it bottom right:} ejected/envelope helium mass fraction.}
\label{fig:MZY}
\end{figure}

\subsection{Nova outbursts -- Model A}
\label{modelAnova}

The initial parameters of Model A are: $ M_{WD} = 1~M_\odot $, $\dot{M}=10^{-11}~M_\odot$~yr$^{-1}$ and $T_{WD}=30\times10^6$~K.
Evolution was followed for $1.5\times10^{10}$~yr (roughly a Hubble time), during which time the WD underwent 1001 nova outbursts.

\begin{figure}
\centering
\includegraphics[width=84mm]{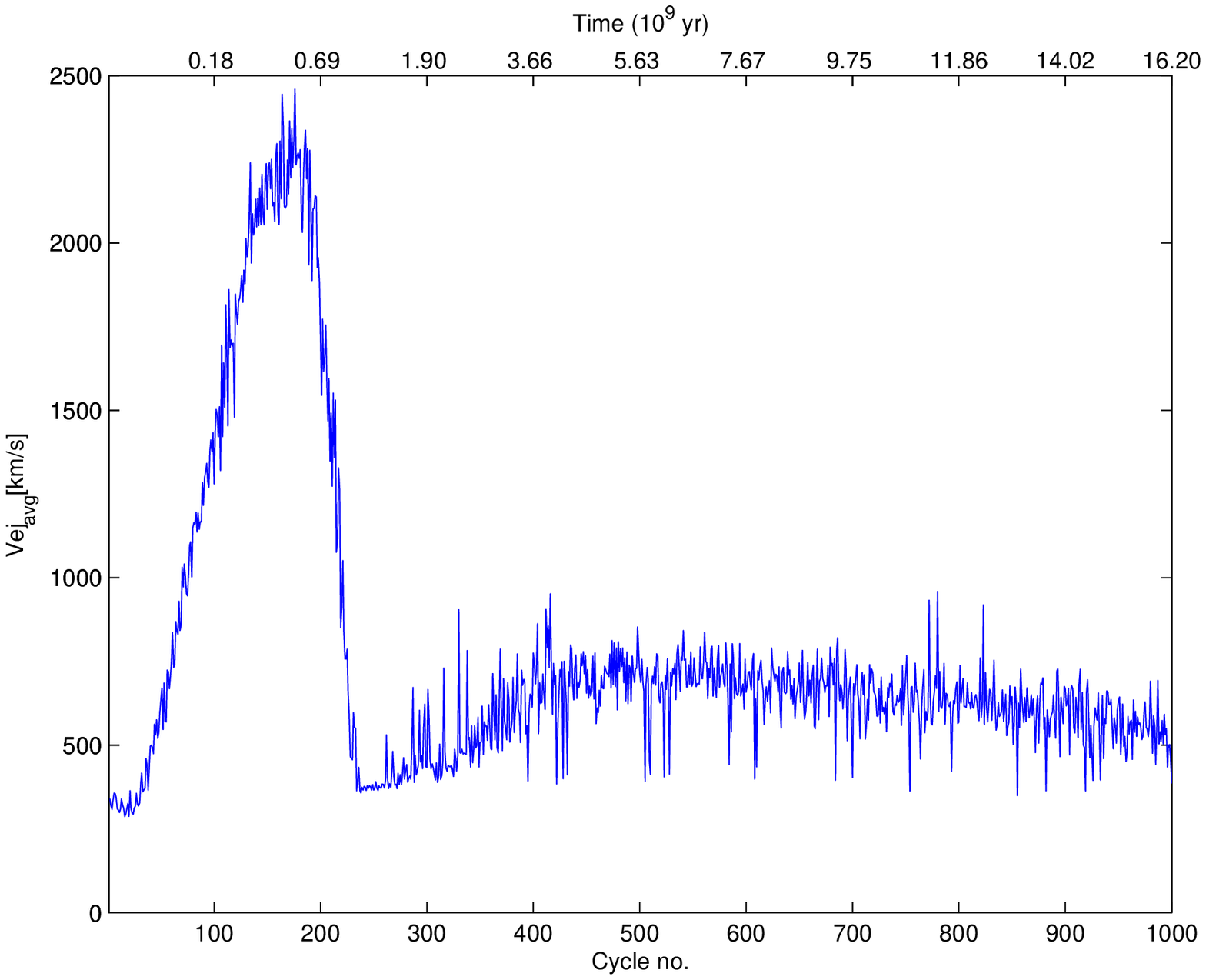}
\includegraphics[width=84mm]{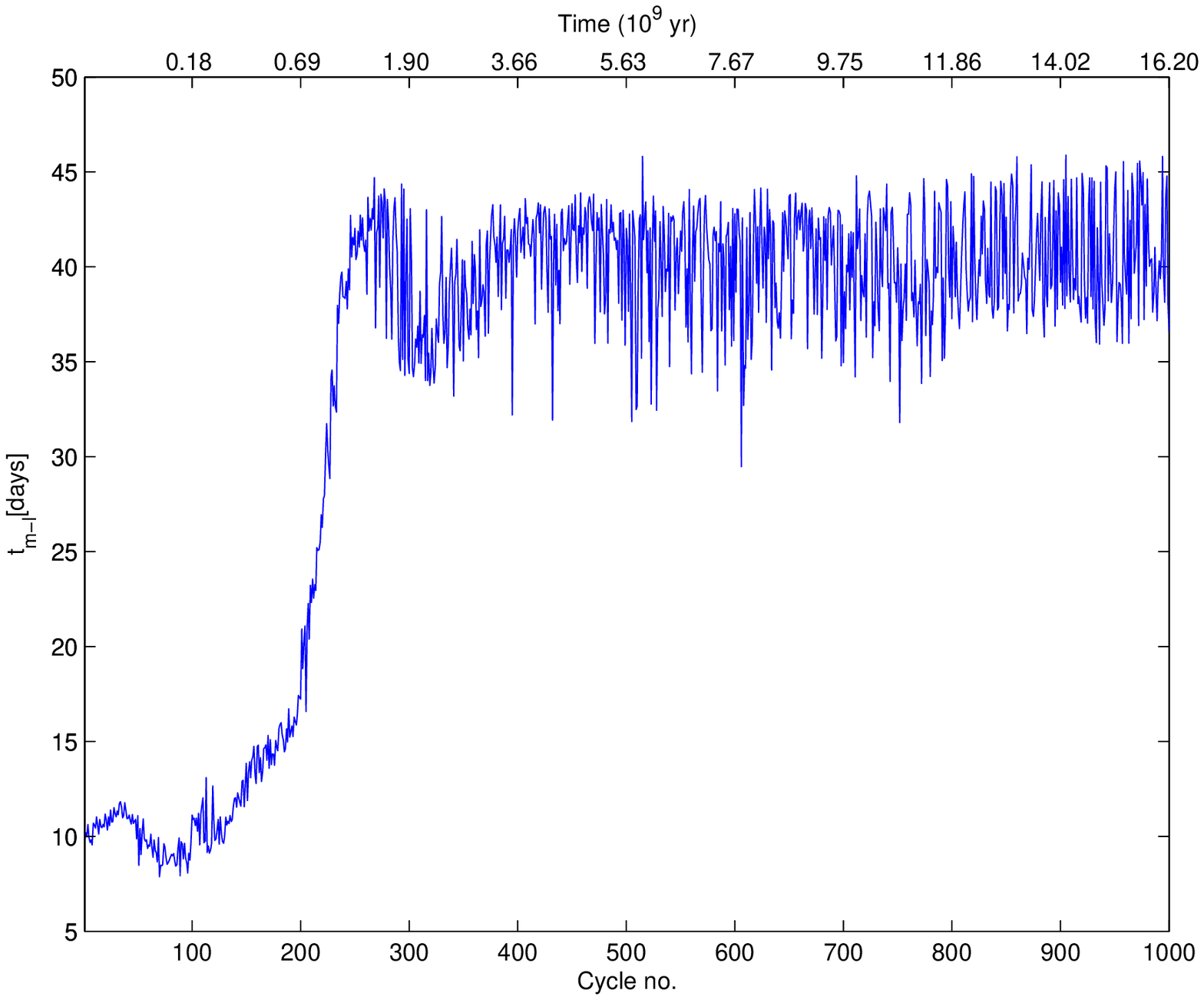}
\caption{Evolution of Model A ($ M_{WD} = 1~M_\odot $ ,$\dot{M}=10^{-11}~M_\odot$~yr$^{-1}$ and  $T_{WD}=30\times10^6$~K): 
{\it top:} average velocity of the ejecta $v_{avg}$; {\it bottom:} duration of the mass loss phase $t_{ml}$.}
\label{fig:Vtml}
\end{figure}

\begin{figure}
\centering
\includegraphics[width=84mm]{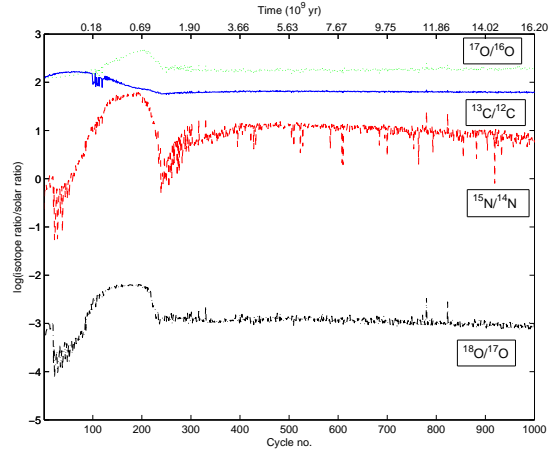}
\caption{Evolution of isotopic ratios normalized by the corresponding solar ratios for Model A: [$^{13}$C/$^{12}$C], 
[$^{15}$N/$^{14}$N], [$^{17}$O/$^{16}$O], [$^{18}$O/$^{17}$O].} 
\label{fig:comp}
\end{figure}

The long-term evolution of the main physical characteristics related to the outbursts that occur on top this accreting WD
is shown in Figures~\ref{fig:MZY}-\ref{fig:comp}. 
The top left 
panel of Fig.~\ref{fig:MZY} shows the
accreted mass $m_{acc}$, ejected mass $m_{ej}$ and the mass of the convective envelope that develops after hydrogen ignition $m_{env}$.
All three increase with time for the first $\sim$240 cycles, or $\sim10^9$~yr, but stabilize thereafter to a very slow rate
of growth. At first, the ejected mass is almost identical to the envelope mass, which means that the remnant, hydrogen-rich layer 
at the end of the outburst is very small, and thus the time of decline of the bolometric magnitude is short as well. We recall that 
the decline starts when the remnant hydrogen is exhausted. Gradually, however, the trend shifts towards the envelope mass being
almost equal to the accreted mass, and the ejected mass being larger than both. This means that WD material is ejected directly
during the late stages of mass loss, without being previously mixed with the accreted material. Observationally, this should be
reflected by a change in composition of the ejecta, which may be detectable if/when the material ejected earlier becomes transparent.
The reason for this effect is a change within the evolution of the nova outburst. When most of the mass contained in the
convective envelope has been ejected, the star contracts back to almost WD size. The high (bolometric) luminosity persists for 
a while longer, until the hydrogen in the envelope remnant is burnt out. When the remnant is massive, however, contraction followed
by heating of the hydrogen-rich remnant layer may ignite it explosively again, which will lead to a second mass loss episode, so
that the entire original envelope and even some additional WD material will be ejected. Such behaviour has already been 
encountered and discussed in previous nova outburst studies (e.g., Prialnik and Livio 1995).   

For each outburst, we calculate the {\it average} abundance of any element $n$ in the ejecta $X_{n,ej}$ 
by 
\begin{equation}
X_{n,ej}=\int \dot M_{m-l}X_{n,S}(t) dt/m_{ej},
\label{abundance}
\end{equation}
where $X_{n,S}(t)$ is the respective 
abundance in the outermost mass shell from which mass is removed (see Section~\ref{method})
at a given time; it is not necessarily
constant (see also Kovetz and Prialnik 1997), as it depends upon the mixing history of the envelope following 
the thermonuclear runaway. The integration is carried over the entire mass loss phase of the cycle.
The ejecta average metallicity $Z_{ej}$ decreases with repeated cycles, as the accreted mass increases. At the same time the helium 
mass fraction $Y_{ej}$ increases, but both reach a plateau after 240 cycles, when
the accreted and ejected masses stabilize as well. At this stage, the breakdown of $Z_{ej}$ into the most abundant elements is:
N - 0.09, O - 0.08, and C - 0.04. 
The difference between $Z_{ej}$ and $Z_{env}$ is a direct consequence of the
relationship between $m_{ej}$ and $m_{env}$, explained above: $Z_{ej} > Z_{env}$, when during the secondary mass loss episode,
WD material ($Z\approx1$) that was not previously mixed with the accreted material, is ejected as well. 

%
The maximal temperature attained at outburst reaches a maximum of about $2.3\times10^8$~K
around the 400'th cycle and declines very slowly thereafter.

Perhaps the most interesting result of this long-term calculation is the relatively sharp transition from fast 
nova outbursts --- for roughly the first 200 cycles --- to moderately fast nova outbursts from the 250'th outburst onwards, as
shown by the duration of the mass loss episode plotted in the lower panel of Fig.~\ref{fig:Vtml}. Even more intriguing is
the behaviour of the average expansion velocity, shown in the upper panel: as the fast nova outbursts progress, the velocity 
goes through a maximum and
settles to an almost constant value of about 500~km~s$^{-1}$ for the moderately fast outbursts. It thus appears
that fast novae may exhibit a range of expansion velocities ranging between 500 and 2500~km~s$^{-1}$, while slower ones
should have typically lower velocities. We should bear in mind, however, that a low average velocity may still allow much
higher peak velocities for short periods of time. Of course, these conclusions relate to outbursts obtained for a 1~$M_\odot$ WD and
may change with the WD mass. 
 
The isotopic ratios shown in Fig.~\ref{fig:comp} undergo a similar evolution: a changing, non-monotonic pattern along the
first 250 fast nova cycles, settling down to constant ratios for most of the evolution. We note significant overabundances
of $^{13}$C and $^{17}$O, compared to solar abundances (taken from Lodders, 2003), 
a somewhat lower overabundance of $^{15}$N, while $^{18}$O is underabundant.
We also note that for the first few tens of outbursts the trends of change are different and variable. 
Thus, in this respect, results based
on one or even a few nova cycles may be misleading. In other respects, however, a general trend is already apparent and maintained
from the beginning.

\subsection{Nova outbursts -- Model B}
\label{modelBnova}

\begin{figure}
\centering
\includegraphics[width=90mm]{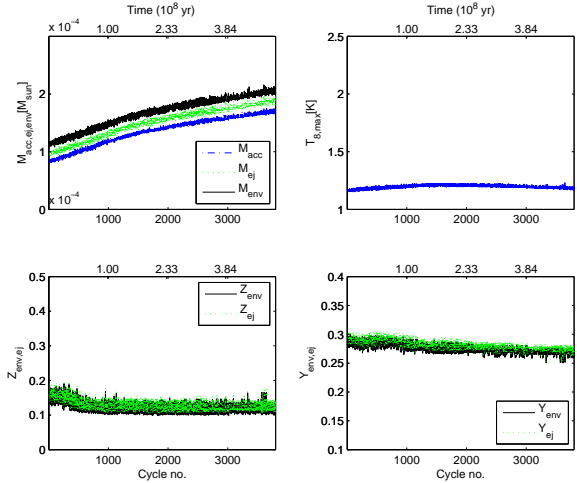}
\caption{Evolution of Model B ($ M_{WD} = 0.65~M_\odot $, $\dot{M}=10^{-9} M_\odot$~yr$^{-1}$ and  $T_{WD}=50\times10^6$~K): 
{\it top left:} accreted/ejected/envelope mass; {\it top right:} maximal temperature attained at outburst; 
{\it bottom left:} ejected/envelope metallicity; {\it bottom right:} ejected/envelope helium mass fraction.}
\label{fig:MZY065}
\end{figure}

The initial parameters of Model B are: $ M_{WD} = 0.65~M_\odot $, $\dot{M}=10^{-9}~M_\odot$~yr$^{-1}$ and $T_{WD}=50\times10^6$~K.
Evolution was followed for a much shorter time, $\sim5\times10^{8}$~yr, but during this time the WD underwent over
3000 nova outbursts. The total mass transferred to it by accretion amounted to $\sim0.5~M_\odot$, which is of the order
of a typical red dwarf binary companion mass. Thus, in spite of the relatively short evolution time, the companion mass may
have been exhausted by its end. 

\begin{figure}
\centering
\includegraphics[width=84mm]{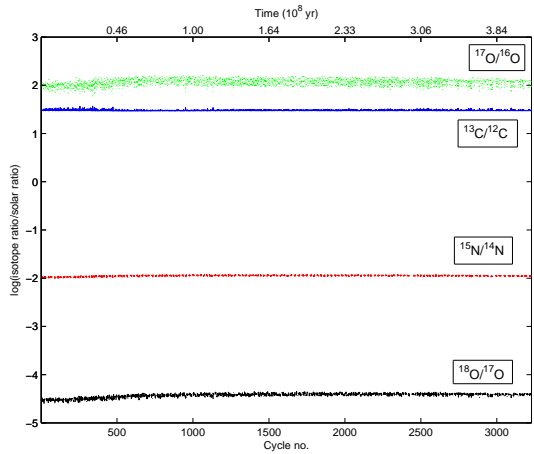}
\caption{Evolution of isotopic ratios normalized by the corresponding solar ratios for Model B: [$^{13}$C/$^{12}$C], 
[$^{15}$N/$^{14}$N], [$^{17}$O/$^{16}$O], [$^{18}$O/$^{17}$O].} 
\label{fig:comp065}
\end{figure}

The long-term evolution of the main physical characteristics related to nova outbursts that occur on top this accreting WD
is shown in Figures~\ref{fig:MZY065} and \ref{fig:comp065}. 
Here, due to the high accretion rate, the evolutionary timescale is much shorter; thus, for example, 1000 cycles which spanned
$\sim 1.5 \times 10^{10} $~yr for Model A, take only $\sim 10 ^ {8} $~yr for model B. The times are not precisely inversely
proportional to the corresponding accretion rates because the accreted masses required for outbursts to be triggered depend on the WD
mass.

The striking difference between Models A and B, which is immediately obvious from the comparison of Figs.~\ref{fig:MZY} and 
\ref{fig:MZY065} as well as Figs.~\ref{fig:comp} and \ref{fig:comp065}, is the constancy (or monotony)
in the evolution of the outburst characteristics for the low-mass WD, 
in contrast to the sharp changes that take place in the course of evolution of the more massive
one. But, in fact, indication for this trend of behaviour was already provided by the grid of nova models (Yaron et al. 2005).
We note, in particular, that for Model B the envelope mass is always larger than both the accreted and the ejected masses.
The mass fractions of the most abundant elements of the ejecta are 0.065 for N, 0.055 for O, and 0.004 for C. 
Regarding isotopic ratios, we note that $^{15}$N is underabundant in the present case, whereas for Model A it was 
overabundant at various levels throughout most of the evolution.
The average expansion velocity oscillates between 150~km~s$^{-1}$ and 220~km~s$^{-1}$, with no particular trend. 
The outburst duration, as measured by the extent of the mass loss phase, increases gradually from $\sim$80 to $\sim$160 days
along the first 1000 cycles and oscillates between these limits thereafter.
 
\subsection{Evolution of the WD -- Models A and B}
\label{modelABWD}

One of the most puzzling questions related to nova outbursts is their effect, if any, on the characteristics of
the underlying WD, keeping in mind that the outbursts are confined to the outermost layers of the WD, less than a thousandth of
its mass. Previous studies, even if they considered several cycles, did not span a sufficient fraction of the WD evolutionary
timescale for answering this question. An indication that the WD core may continue to cool despite the heat generated
by nova outbursts was given by Prialnik (1987). In the present calculations the evolution of the WD was carried out for
$1.5\times10^{10}$~yr (about a Hubble time); we are thus in a position to answer this question.

\begin{figure}
\centering
\includegraphics[width=90mm]{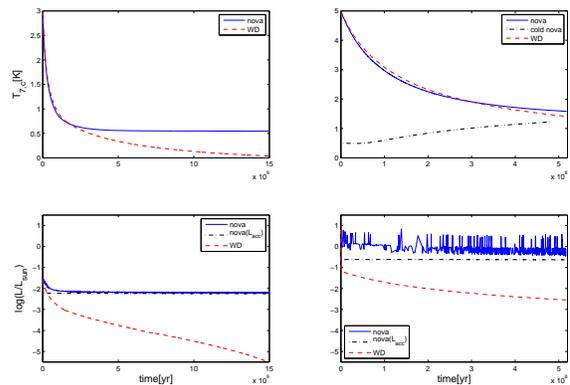}
\caption{Evolution of the central temperature ({\it top}) and of the luminosity ({\it bottom})
of the accreting WD and a cooling WD of the same mass and initial core
temperature: {\it left} -- Model A; {\it right} -- Model B. The accretion
luminosity, eq.~(\ref{eq:Lacc}), is also shown in the lower panels. An additional curve for Model B
shows the long-term evolution through almost 3000 nova cycles, adopting the same mass and accretion rate, 
but a 10 times lower initial temperature, $T_{WD}=5\times10^6$~K (see discussion in Section~\ref{comparison}).}
\label{fig:TL}
\end{figure}

Using the same code, we followed the evolution of the same initial WD configuration, omitting accretion, to
obtain the cooling curve of the WD. The central temperatures of the undisturbed WD and of the accreting one are
compared in Fig.~\ref{fig:TL}. For the 1~$M_\odot$ WD (Model A), the accreting WD cools at the same rate as 
the single one for a period of $\sim10^9$~yr
(equivalent to about 250 cycles), but then the cooling trend slows down and the WD core tends to a constant temperature 
of $\sim5\times10^6$~K. The evolutionary timescale for Model B is short compared to the cooling timescale of a WD, hence
the cooling curves are still close even after several thousand cycles, which amount to only a few $10^8$~yr. 
The luminosities of the cooling WD and of the accreting WD in quiescence (during the accretion
episodes separating outbursts) are compared in Fig.~\ref{fig:TL}. The luminosity of the accreting WD has two sources:
the heat emanated by the cooling WD core and the energy imparted by accretion, known as accretion luminosity, given by
\begin{equation}
\label{eq:Lacc}
L_{acc}=\frac{\alpha_{acc} GM_{WD}\dot{M}}{R_{WD}}
\end{equation}
where $\alpha_{acc}$ is taken to be $0.15$, following Regev and Shara (1989). The outer layers of the WD adjust
so as to re-emit this influx of energy, in a similar manner to an irradiated star (Kovetz et al. 1988). 
We note that only at the beginning, while the WD core is still 
relatively hot, does the core luminosity exceed the accretion luminosity.

\begin{figure}
\centering
\includegraphics[width=90mm]{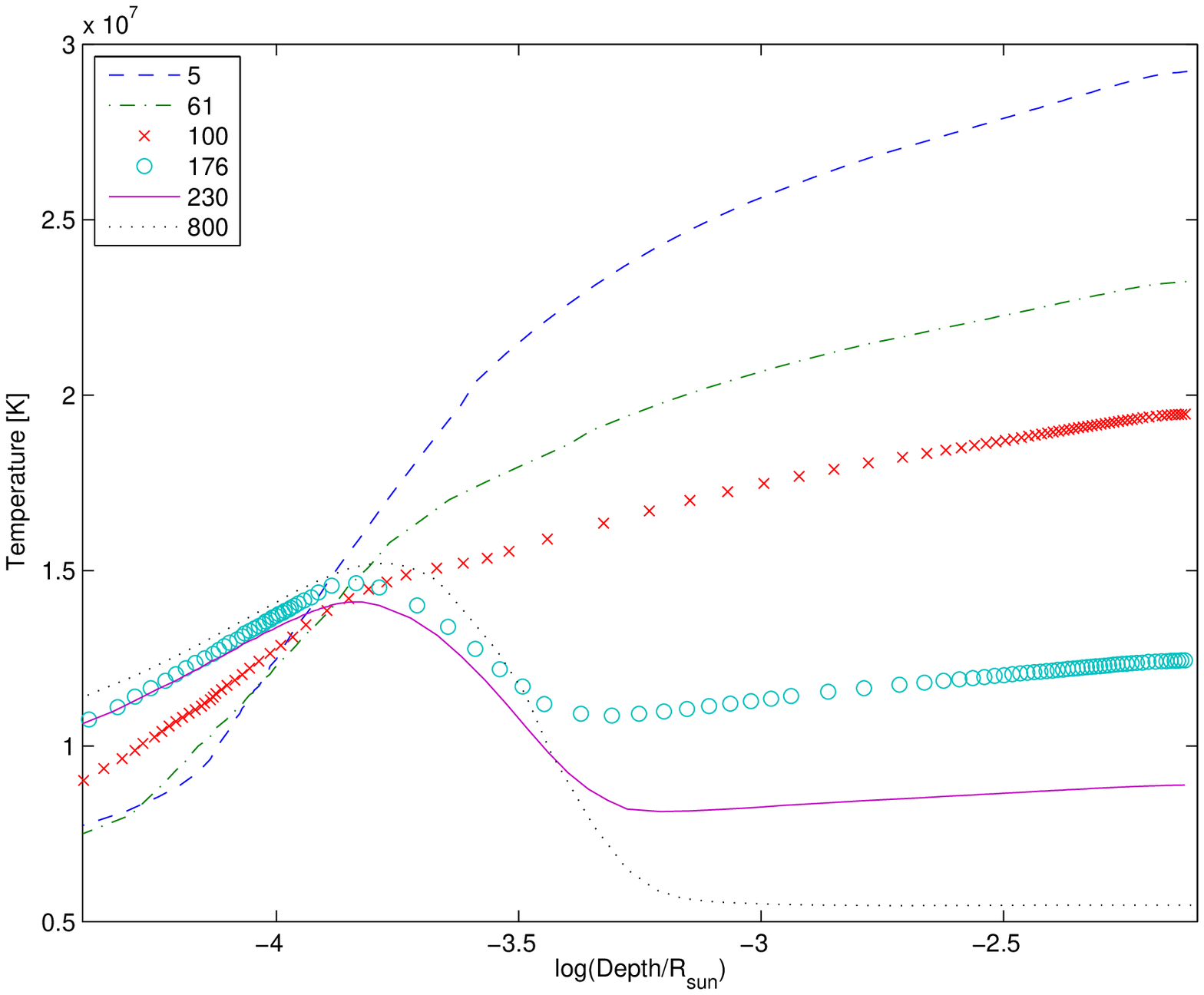}
\includegraphics[width=90mm]{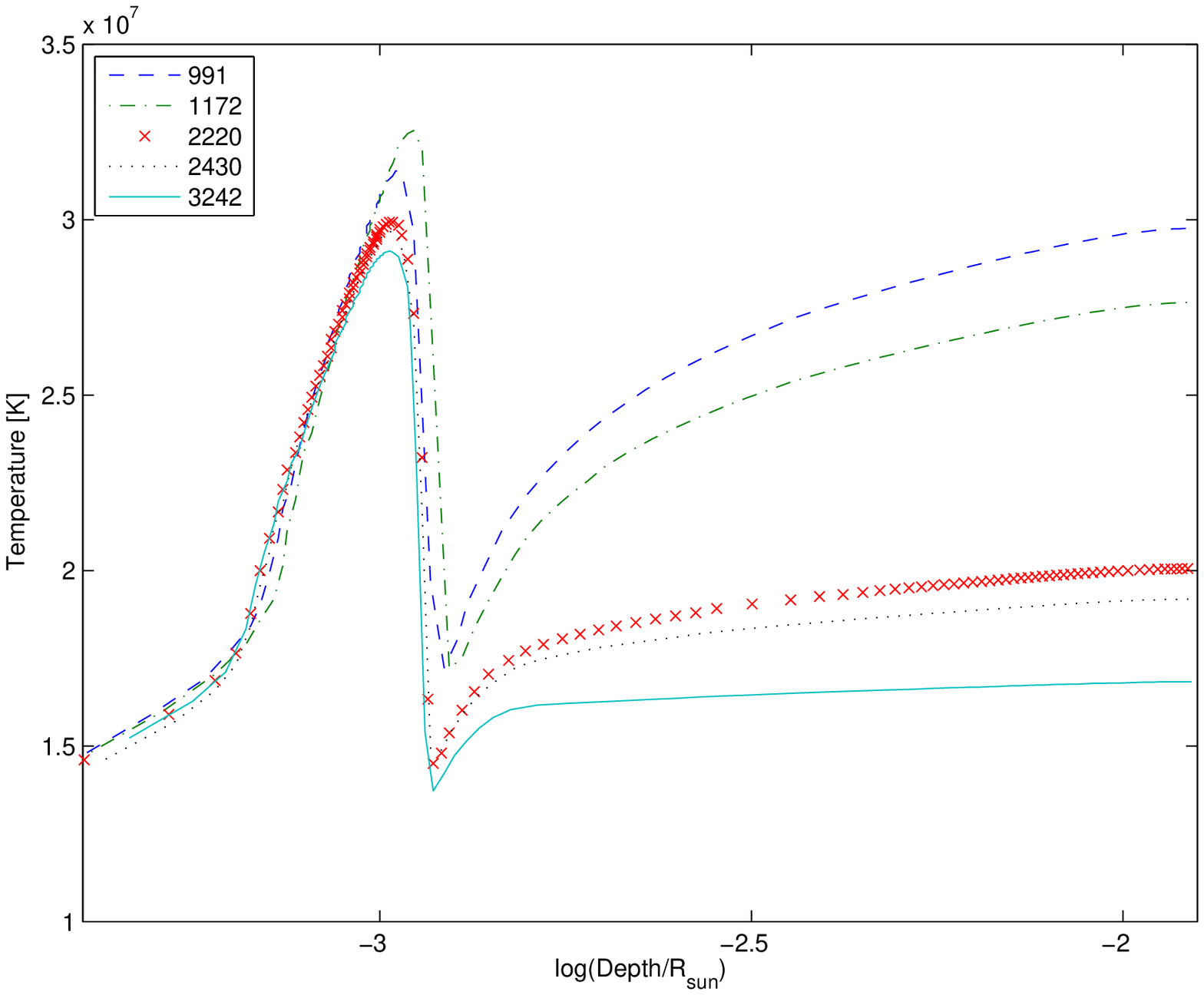}
\caption{Temperature profiles from the surface inwards on a logarithmic scale at the end of a cycle (onset of accretion)
for several cycles --- as marked --- along the evolutionary course of Model A ({\it top}) and of Model B ({\it bottom}). 
luminosity, eq.~(\ref{eq:Lacc}), is also shown:{\it left} -- Model A; {\it right} -- Model B.}
\label{fig:profT}
\end{figure}

In order to understand the thermal evolution of the WD, we plot in Fig.~\ref{fig:profT} temperature profiles at various
times during evolution (marked by the cycle number). All profiles correspond to the end of a nova cycle, just before the
onset of a new accretion episode. Thus the configuration includes the WD and the eventual remnant of the envelope at the
end of the mass loss phase. At first, for several hundred cycles, the temperature gradient is negative everywhere. Gradually,
however, the outer layers of the WD, well beneath the outermost one that is directly affected by the outburst (mainly through
diffusion, which changes its composition), are heated and a temperature inversion develops. Nevertheless, for many additional cycles,
this temperature "wall" does not prevent the core from cooling, since the temperature still decreases between the
centre and the inversion zone. But the slope of the core temperature profile slowly decreases and finally flattens.
At this stage the WD core ceases to cool. It may subsequently start to heat up, but for this to happen, more than a Hubble time
seems to be required for the massive WD, and several thousand additional nova cycles (that is, a massive companion)
for the less massive one. 
In conclusion, the shape and evolution of the temperature profiles explains the difference (or similarity) between the cooling
curves of an accreting and an undisturbed WD, shown in Fig.~\ref{fig:TL}.

\begin{figure}
\centering
\includegraphics[width=90mm]{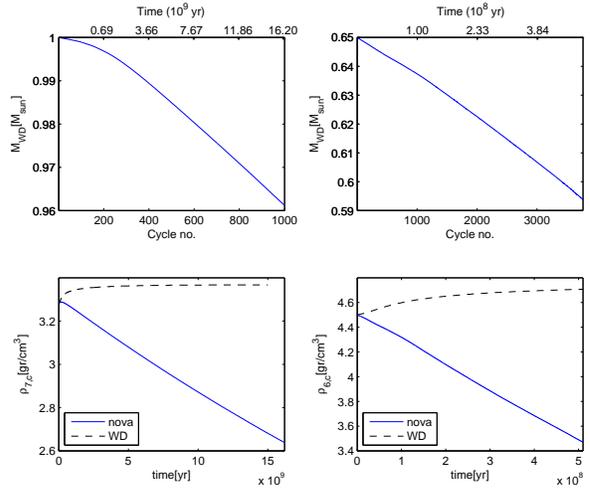}
\caption{Decrease of the WD mass due to mass loss during nova outburst cycles ({\it top}) and
evolution of the central density of the accreting WD and a cooling WD of the same mass and initial core ({\it bottom})
temperature: {\it left} -- Model A; {\it right} -- Model B.}
\label{fig:Mrho}
\end{figure}

The evolution of the central density of the accreting WD is shown in the lower panel of Fig.~\ref{fig:Mrho}, together
with that of the single WD. While the latter increases approaching an asymptotic value, the former --- for Model A --- increases at first
but decreases steadily thereafter. The reason for this behaviour lies in the competition between the already
mentioned cooling of the core, which drives the density up, and decrease of the WD mass (shown in the upper panels
of Fig.~\ref{fig:Mrho}), which drives it down.
Applying to the WD centre the equation of state for degenerate matter to first order in temperature (e.g., Landau and Lifshitz, 1969), 
we have 
\begin{equation}
\label{eq:steq}
P_c=\beta\rho_c^{5/3}\left(1+\alpha \frac{T^2_c}{\rho_c^{4/3}}\right)
\end{equation}
where, in standard notation, 
\begin{equation}
\label{eq:betalfa}
\beta=\frac{\hbar^2}{5m_e} \frac{(3 \pi ^2)^{2/3}}{(m_H \mu_e)^{5/3}} \qquad 
\alpha=\frac {5m^2_e k^2}{6 \hbar^4}\frac{ (m_H \mu_e)^{4/3}}{ (3 \pi ^2)^{1/3}}
\end{equation}
and $ \mu_e \approx 2 $. 

On the other hand, hydrostatic equilibrium for a polytropic equation of state requires
\begin{equation}
\label{eq:hyeq}
P_c=(4 \pi)^{1/3} B_n G M^{2/3}_{WD} (\rho_c)^{4/3}
\end{equation}
where $ B_n$ is a constant determined by the polytropic index $n$ (e.g., Chandrasekhar, 1967), thus $B_{1.5} = 0.206 $
for $P\propto\rho^{5/3}$.
Combining eqs.~\ref{eq:steq} and
\ref{eq:hyeq}, we obtain a relation between $\rho_c$, $T_c$ and $M_{WD}$, of the form
\begin{equation}
\label{eq:rhoTM}
\beta \rho_c^{1/3} + \alpha \beta \frac{T_c^2}{\rho_c}= \gamma M_{WD}^{2/3}
\end{equation}
where
\begin{equation}
\label{eq:gamma}
\gamma=(4 \pi)^{1/3} 0.206 G 
\end{equation}
Taking the time derivative of eq.~(\ref{eq:rhoTM}), we obtain
\begin{equation}
\label{eq:ddt}
\frac{\beta}{2 \rho_c}\left(\frac{\rho_c^{1/3}}{3} - \frac{\alpha T_c^2}{\rho_c}\right) \frac{d\rho_c}{dt} = 
- \alpha \beta \frac{T_c}{\rho_c} \frac{dT_c}{dt} + \frac{\gamma}{3} M_{WD}^{-1/3} \frac{dM_{WD}}{dt}
\end{equation}

\begin{figure}
\centering
\centerline{\includegraphics[width=45mm]{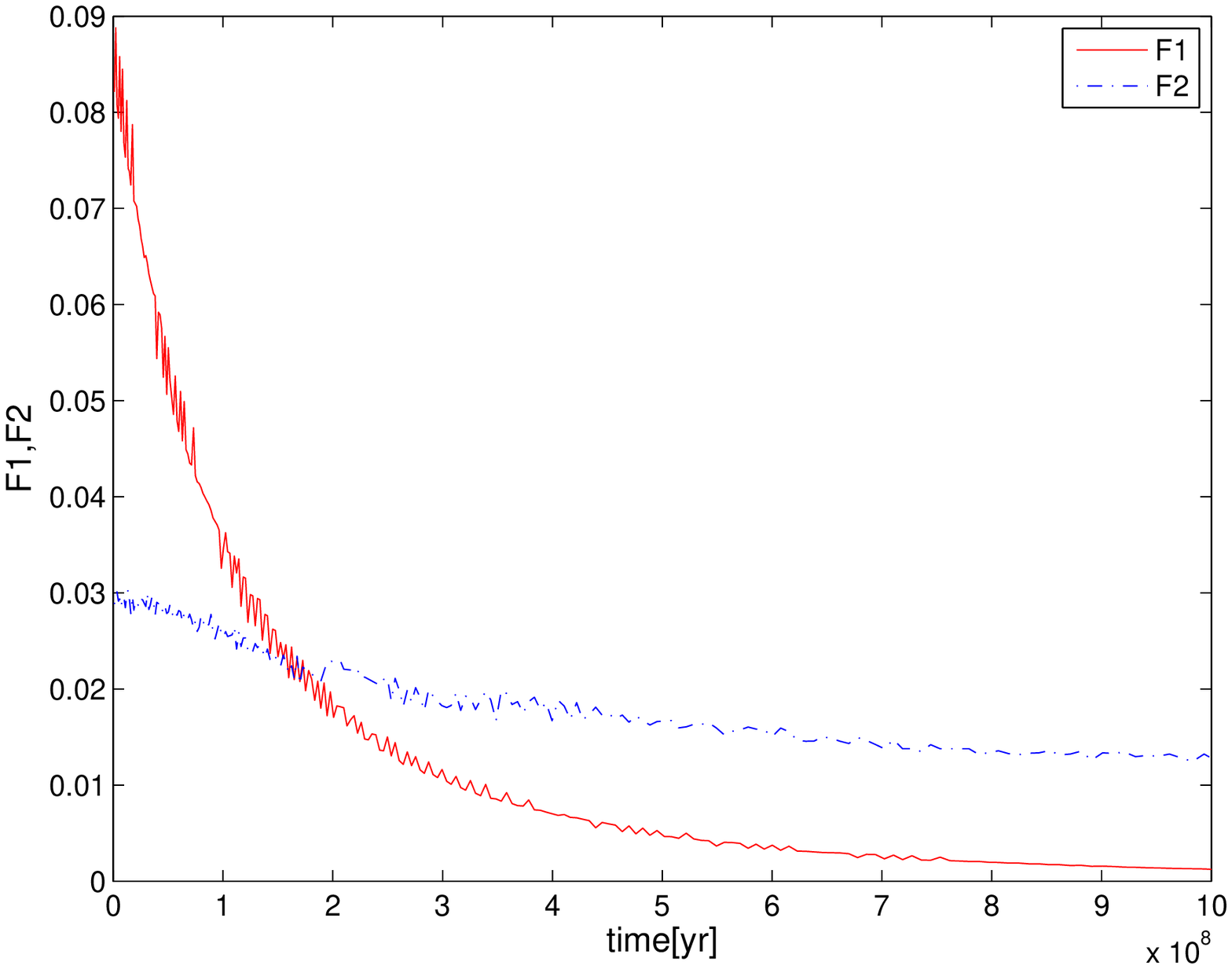}\ \includegraphics[width=45mm]{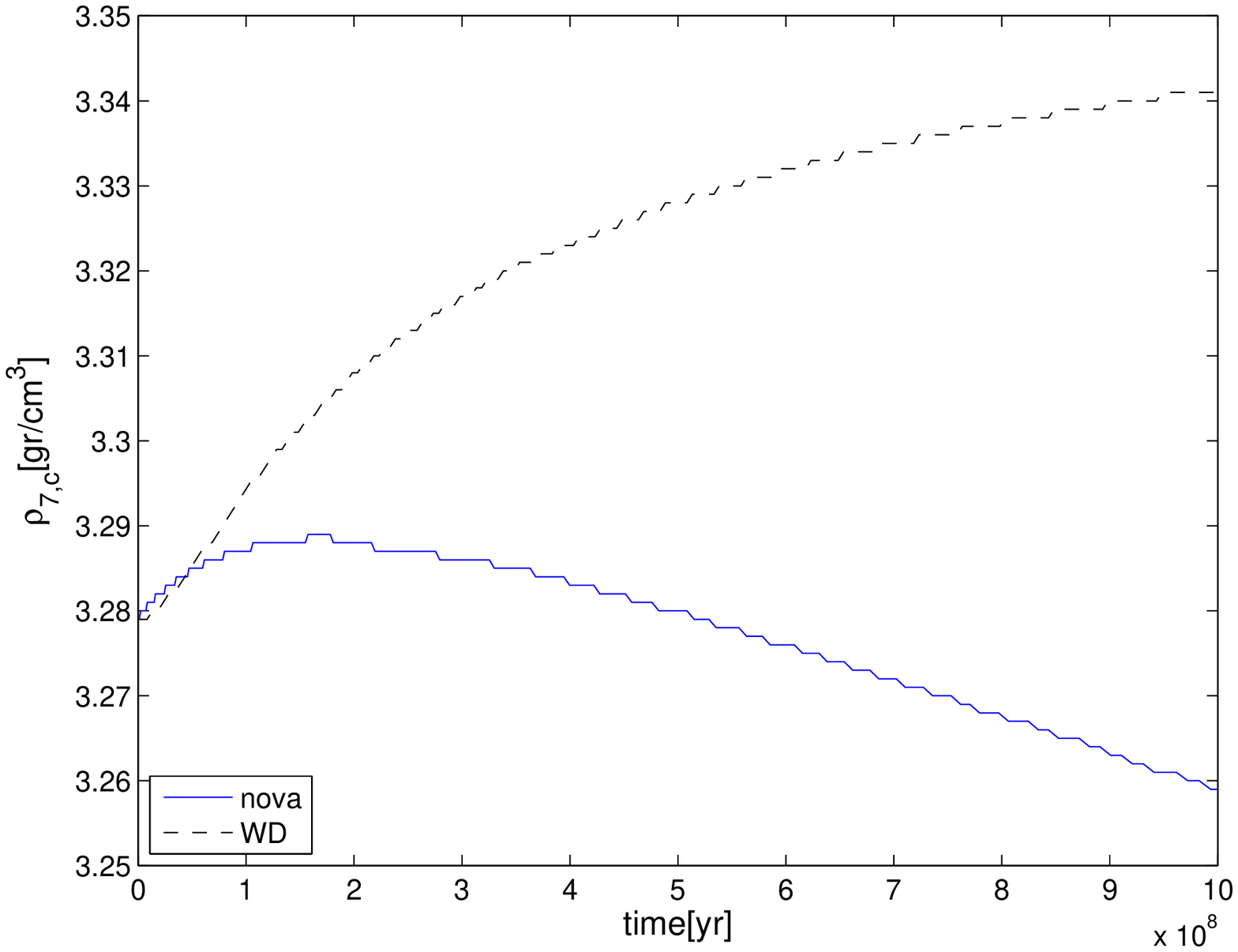}}
\caption{{\it Left:} Evolution of the LHS terms of eq.~(\ref{eq:ddt}) for Model A (see text for details):
$F1$ represents the first term, which includes the derivative of temperature, and $F2$ - the second
term, which includes the derivative of the WD mass. {\it Right:} Zoom in into the evolution of the
central density (shown in Fig.~\ref{fig:Mrho}), to show the rise and decline of $\rho_c$ during 
early evolution.}
\label{fig:Prho}
\end{figure}

Since $\alpha$ is a very small number, deriving from the small correction to the degenerate
equation of state due to temperature, the coefficient of the density derivative on the LHS of
eq.~(\ref{eq:ddt}) is positive, regardless of the changes with time in the temperature and density
values. The terms on the RHS have opposite signs, since both $dT_c/dt$ and $dM_{WD}/dt$ are
negative. Since $T_c$ changes rapidly with time at the beginning and very slowly thereafter, while
the WD mass decreases at an almost constant rate, the LHS will change sign from positive to
negative at some point and thus the density will go through a maximum. The competition between
the two terms is shown in Fig.~\ref{fig:Prho} for Model A. A correction factor was introduced in the second term,
to take account of the fact that for a 1~$M_\odot$ WD relativistic effects on the equation of state
~(\ref{eq:steq}) are non-negligible. The density maximum corresponds to the intersection.
 
However, since the rate of change of the WD mass is roughly proportional to the rate of accretion,
\begin{equation}
\label{eq:mdotwd}
\frac{dM_{WD}}{dt}\approx - \dot M\ \frac{m_{ej}-m_{acc}}{m_{acc}}\ ,
\end{equation}
for a very high accretion rate the second term on the RHS of eq.~(\ref{eq:ddt}) will always dominate and the
central density will decrease monotonically. This is the situation for Model B, where the accretion rate
is two orders of magnitude higher than for Model A. Equally, if $dT_c/dt > 0$, the
central density will decrease monotonically.

\subsection{Energy budget}
\label{energy}

\begin{figure}
\centering
\includegraphics[width=84mm]{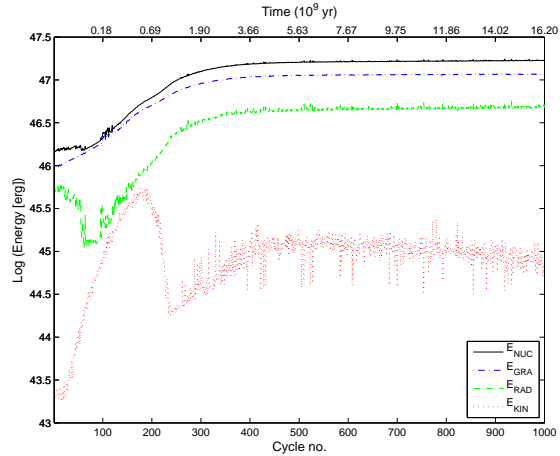}
\caption{Energy sources and sinks as a function of time: total nuclear energy released during a nova cycle,
gravitational binding energy of the ejecta, total radiated energy during outburst and
kinetic energy of the ejecta, for Model A.}
\label{fig:E}
\end{figure}

It is instructive to examine the energy budget of nova outbursts and its eventual change with time. The nuclear
energy generated during the thermonuclear runaway and during the equilibrium burning of the remnant hydrogen
(if any), $E_{nuc}$, is spent in radiation at close to (or exceeding) Eddington luminosity during the outburst, 
$E_{rad}$, kinetic energy of the ejecta, $E_{kin}$, and 
gravitational energy required to lift the ejected material out of the potential well of the WD, $E_{grav}$. 
These are plotted in Fig.~\ref{fig:E} for Model A.
The last term is not computed during calculations, but estimated by $E_{grav}\approx M_{WD}m_{ej}/R(M_{WD})$,
which is an upper limit to the actual value, since $R(M_{WD})$ is the radius at the lower boundary of the ejected
mass.
We note that the bulk of nuclear energy is spent in work against the gravitational field of the WD. We also note
that the radiated energy exceeds the kinetic energy and, for most of the evolution, quite significantly.
Finally, the sum of sinks is lower than the source, that is,
\begin{equation}
E_{nuc}\ \ga\ E_{grav}+E_{rad}+E_{kin}
\end{equation}
which means that a small fraction of the heat generated at outburst is conducted into the WD. 
We may estimate the total thermal energy that the WD gains for the entire evolution time, or more precisely,
a lower limit for it, by 
$\sum_{cycles}E_{nuc}-(E_{grav}+E_{rad}+E_{kin})\ga2\times10^{48}$~erg. 
This energy represents about 1\% of the total energy of the WD and 
explains the
very slow change in slope of the core temperature profile shown in Fig.~\ref{fig:profT}.
During evolution, the WD expands due to mass loss (see Fig.~\ref{fig:Mrho}), and therefore its gravitational potential
(binding) energy --- as computed by the evolution code --- increases (that is, becomes less negative), while the
internal energy (mostly the energy of the degenerate electron gas) decreases by roughly the same amount ($\sim15$\%),
as expected (cf. Mestel and Ruderman 1967).
 
For the 0.65~$M_\odot$ WD, the budget is slightly different. All energy values are almost constant in time, as are
most of the other nova characteristics. Typical values are: $E_{nuc}\approx7.6\times10^{46}$~erg, 
$E_{grav}\approx2.7\times10^{46}$~erg, $E_{rad}\approx4.2\times10^{46}$~erg, and finally, $E_{kin}\approx8\times10^{43}$~erg.
We note that in this case the radiated energy is the major energy sink --- due to the long duration of the outburst --- while
the total kinetic energy is only a very small fraction of the nuclear energy generated, considerably lower than the
radiated energy.

\section{Discussion and conclusions}
\label{conclusions}

\subsection{Comparison with earlier studies}
\label{comparison}

Model A starts at  a core temperature of $30\times10^6$~K, which decreases during the WD's evolution. After 210 cycles 
($7.8\times10^8$ ~yr), the temperature reaches $10\times10^6$~K. This result enables a comparison with the grid of 
nova models calculated by Prialnik and Kovetz (1995) and Yaron et al. (2005), where the different values of WD core
temperature were set as initial independent parameters. Similarly, the 0.65~$M_\odot$ WD, which started at a core temperature of 
$50\times10^6$~K, reaches $30\times10^6$~K after 970 cycles ($9.7\times10^7$ ~yr). 
The results are compared in Table~\ref{tab:compar}. We note a striking similarity between the results, allowing for the fact that
the WD mass changes as well along with the core temperature. In the case of the 0.65~$M_\odot$ WD, for example,
it dropped to 0.637~$M_\odot$ by the time the core temperature dropped to $30\times10^6$~K.

\begin{table*}
\centering
 \caption{Comparison between present calculations and the grid of nova models (Yaron et al., 2005) for $M_{WD}=1~M_\odot$ and
$\dot M=10^{-11}M_\odot$~yr$^{-1}$ and for $M_{WD}=0.65~M_\odot$ and $\dot M=10^{-9}M_\odot$~yr$^{-1}$ - Characteristics of 
the nova outbursts.}
\label{tab:compar}
 \begin{tabular}{@{}lccccccccc@{}}
\hline
 $        $ & $m_{acc}$ & $m_{ej}$ & $Y_{env}$ &
 $Y_{ej}$ & $Z_{env}$ & $Z_{ej}$ & $T_{8,max}$ & $v_{avg}$ & $t_{m-l}$\\
 $     $ & $(M_\odot)$ &
 $(M_\odot)$ & $    $ & $   $ & $   $ & $   $ &$(10^8~\degr K)$ &$(km/s)$ &$(days)$\\
  \hline
 Model A -- Grid & $8.72E-05$ & $1.00E-04$ & $0.2475$ & $0.2730$ & $0.1529$ & $0.1917$ & $2.09$ & $1063$ & $22.3$\\
 Model A -- Evolution & $8.77E-05$ & $1.15E-04$ & $0.2418$ & $0.2480$ & $0.1730$ & $0.2665$ & $2.11$ & $1349$ & $35.4$\\
\hline
 Model B -- Grid & $1.11E-04$ & $1.21E-04$ & $0.2884$ & $0.2956$ & $0.1037$ & $0.1102$ & $1.21$ & $156$ & $679$\\
 Model B -- Evolution & $1.17E-04$ & $1.33E-04$ & $0.2831$ & $0.2887$ & $0.1248$ & $0.1322$ & $1.20$ & $206$ & $106$\\
\hline
 \end{tabular}
\end{table*}

Townsley and Bildsten (2004) studied the problem of accreting WDs analytically, {\it assuming} that the WD core reaches
an equilibrium temperature and this is achieved by heating of the core due to release of gravitational energy during compression
of the accreted material as well as quiet nuclear burning during the nova cycle. Their study, by its nature,
does not take into consideration evolutionary effects, such as the decrease of the WD mass, and dynamical phases
of the outburst, which determine to a large extent the energy budget. Our full-scale calculations show that, indeed, a steady
state (or equilibrium) is eventually attained or approached asymptotically, but on a {\it very long} timescale. 

The equilibrium core temperatures that we find
for the two cases considered here are considerably higher than those derived by Townsley and Bildsten (loc. cit.); the accreted masses,
equivalent to their ignition masses, for the same cases, are similar. In order to test whether the steady state temperature
is, indeed, an equilibrium temperature, we have run a calculation similar to Model B, but for a very low initial WD temperature,
lower than the asymptotic value. The evolution was again followed for close to 3000 full nova cycles and we found the WD temperature
to rise slowly towards the same steady state value of about $1.5\times10^7$~K, as the cooling, initially hot, WD. We wish to stress, 
however, that in both cases the asymptotic value is approached very slowly compared with the evolutionary timescale of the nova 
binary system, on which the mass ratio of the components changes very significantly, and the mass transfer rate is expected to 
change as well. Consequently, the equilibrium core temperature may not be assumed to occur {\it a priori}.

\subsection{Summary}
\label{summary}

We have carried out long-term continuous evolutionary calculations of accreting WDs through thousands of nova cycles, 
where each cycle was followed in detail through its different phases: accretion, thermonuclear runaway, expansion, mass loss
and contraction. 

Ths study shows, for the first time, the long-term effect nova outbursts have on the structure and evolution of the WD at
the surface of which they occur. At the same time, it shows the effect of the changing WD characteristics, in turn, on the
nova outbursts. We have chosen two very different combinations of WD mass and accretion rate in order to test these two
types of effect. The main results and conclusions reached may be summarized as follows:

\begin{enumerate}
\item
 The WD physical parameters change considerably with time in both cases, in particular:
  \begin{enumerate}
  \item
The WD cools as a single, undisturbed WD for a while, but after a time, cooling slows down and will eventually cease. The maximum 
temperature attained in the burning shell $ T_{max} $ increases at first, but settles at an almost constant value as cooling
of the WD slows down. 
  \item
 The WD luminosity decreases with time, but in contrast to a single WD, only down to the accretion luminosity value, where it remains 
constant. This, of course, applies only to the quiescent phases of accretion between outbursts.
  \item
The WD mass decreases steadily for the cases considered, but not by much. The mass of the donor star, by contrast,
decreases considerably: by 0.16~$M_\odot$ for Model A, and by 0.51~$M_\odot$ for Model B. 
  \item
 The WD becomes less dense  as its mass decreases, as expected from the relation $\rho_c(M)$ characteristic of degenerate stars.
  \end{enumerate}
\item
On the long-term scale, the outburst characteristics change considerably for the massive WD (Model A), but 
only slightly for the low-mass one (Model B). In particular:
  \begin{enumerate}
  \item  
The accreted and ejected masses increase with time, as the WD temperature decreases. 
  \item
The heavy element abundance ($Z$) decreases with time for Model A, while it oscillates around a constant value for model B. 
  \item
The helium mass fraction increases with time for Model A, while again, it oscillates around a constant value for Model B. 
  \item
Isotopic ratios settle eventually into constant values for Model A and are practically constant throughout for Model B.  
In both cases $^{13}$C and $^{17}$O are overabundant as compared to solar values, while $^{18}$O is underabundant; $^{15}$N,
however, is overabundant in one case (A) and underabundant in the other (B).
  \end{enumerate}
\item
The changes in physical characteristics are not necessarily monotonic; some, such as the expansion velocity, go through an extremum. 
Consequently, one should be careful when using parametrised grids of models and interpolating between them. Nevertheless, the
agreement between the long-term calculations and the parametrised models is quite good for many of the nova characteristics.
\item
Both models settle, eventually, into almost steady state where both the WD properties and the outburst characteristics remain
almost unchanged with many repeated cycles. However, this state is reached after hundreds of cycles, which means a long
period of time, or a considerable change in the mass of the donor star. Although our calculations assume a constant
accretion rate throughout, it is to be expected that spurious dynamical effects on the binary orbit that are bound to occur
on a long timescale, or the change in donor mass, or both, will affect the mass transfer rate significantly. A different accretion rate
will tend to a different steady state. It is thus possible that, in reality, the system will never be able to achieve steady state.
\item
For the first time in numerical modelling of nova outbursts, it is possible to estimate the accuracy (or reliability) of 
the results. The numerical "noise" that appears in the various plots may be statistically interpreted as error bars.
\end{enumerate}
 
Of the 3 independent nova parameters, only 2 are left --- the WD mass and the accretion rate --- while the 
third --- the WD temperature --- is determined by the
evolutionary course of the accreting WD in the nova system. The present calculations were based on the arbitrary assumption 
that the accretion rate is constant (and prescribed). This assumption is not realistic; the evolutionary course of the
nova system, taking into account the interaction between its components, determines the mass transfer rate and its evolution,
and future studies should follow the evolution of both members of the binary system consistently.
This leaves the WD mass as the single truly independent parameter of nova outburst evolution.

\section{Acknowledgments}

We are grateful to Mike Shara for refusing to accept that such a calculation as presented here should be impossible to perform.
We wish to thank an anonymous referee for helpful suggestions and comments.

\label{lastpage}

\end{document}